\documentclass[twocolumn,a4paper,reprint,amssymb,aps,prd,groupedaddress,superscriptaddress,nofootinbib]{revtex4-1}
\usepackage[dvips]{graphicx}
\usepackage{amssymb}
\usepackage{filecontents}
\usepackage{amsmath}
\usepackage{color}
\usepackage{float}
\usepackage{textcomp}
\usepackage{bm}
\usepackage{lipsum}
\allowdisplaybreaks
\usepackage{mathtools}
\RequirePackage[colorlinks,citecolor=blue,urlcolor=blue,linkcolor=blue]{hyperref}
\usepackage{yfonts}

\usepackage{xcolor}

\begin{document}

\title {Dynamics of the $N$-body system in energy-momentum squared gravity: Equations of motion
to the first post-Newtonian order}

\author{Elham Nazari}
\email{elham.nazari@mail.um.ac.ir}
\affiliation{Department of Physics, Faculty of Science, Ferdowsi University of Mashhad, P.O. Box 1436, Mashhad, Iran}

\begin{abstract}

In the energy-momentum squared gravity (EMSG), the matter energy-momentum tensor is not conserved due to nonminimal interaction between the usual and modified matter fields.
For this reason, the $N$-body acceleration may host the EMSG effects that can be probed at the solar scale by the perihelion shift of the planets and experimental tests of the Strong Equivalence Principle (SEP). 
To shed light on it, in this paper, we present the $N$-body equations of motion in the weak-field limit of the EMSG theory. To do so, the post-Newtonian (PN) hydrodynamic equations, the viral identities, as well as the corresponding equilibrium conditions are introduced in this theory.
Armed with these relations, we derive the dynamics of the $N$-body system and its PN inter-body metric.
It is shown that the EMSG theory is not ruled out by the classical test, the perihelion advance of Mercury, and the test of SEP. In other words, in the first PN order, it is not possible to constrain the free parameter of this theory and even distinguish it from general relativity (GR) using these local tests.
\end{abstract}
\maketitle

\section{Introduction}

In the history of gravity, since the late 1800s, there have always been attempts to modify the theory of gravity with various theoretical and observational motivations. In the ``beyond-Einstein" period\footnote{ The ``beyond-Einstein" terminology is adapted from \cite{will2018theory}.}, beginning in the middle 1980s, some alternative gravity theories have been devised to solve problems of the standard model of cosmology, known as the Lambda cold dark matter ($\Lambda$CDM) based on general relativity (GR), such as the lack of explanation of the cosmic dark sector, the cosmological constant $\Lambda$ problem \cite{1989RvMP...61....1W,2003RvMP...75..559P,2003PhR...380..235P}, and several tensions between constraints derived from different observations \cite{2021APh...13102606D,2021APh...13102605D,2021APh...13102604D,2021APh...13102607D,2021CQGra..38o3001D}. Other alternative gravity theories have been motivated by particle physics and quantum gravity. For a review on modified gravity theories see \cite{2007IJGMM..04..115N,2010LRR....13....3D,2010RvMP...82..451S,2011PhR...505...59N,2011PhR...509..167C,2015CQGra..32x3001B,2017PhR...692....1N,2023PhR..1034....1D}. 

Among the modified theories of gravity most of which focus on the generalization of the linear function of the Ricci scalar curvature $R$ in the Einstein-Hilbert (EH) action, a new type of modified theory recently has been invented that modifies the matter portion of the standard action of GR. Several models have been proposed for this matter-type modified theory of gravity. In some models, the matter Lagrangian density $\mathcal{L}_{\text{m}}$ is changed to an arbitrary function $f(\mathcal{L}_{\text{m}})$  \cite{2010EPJC...70..373H}. Other models of this theory are constructed by adding a function of the energy-momentum tensor $T^{\mu\nu}$ such as $f(g_{\mu\nu}T^{\mu\nu})$ \cite{2011PhRvD..84b4020H} and $f(T_{\mu\nu}T^{\mu\nu})$ \cite{2014EPJP..129..163K,2016PhRvD..94d4002R,2018PhRvD..97b4011A,2017PhRvD..96l3517B} to the standard action. In the matter-type modified theories of gravity, there is a nonminimal interaction between the usual energy-momentum tensor $T^{\mu\nu}$ and the modified matter field $T^{\mu\nu}_{\text{mod}}$ which is made of different combinations of $\mathcal{L}_{\text{m}}$ in each model \cite{2023arXiv230611717A}.  Therefore, in general, the covariant divergence of the energy-momentum tensor is not necessarily zero, namely $\nabla_{\mu}T^{\mu\nu}\neq0$, and consequently, test particles do not move on geodesics of the curved spacetime, e.g., see  \cite{2010EPJC...70..373H,2016PhRvD..94d4002R,2020PhRvD.102f4016N}. In other words, by introducing auxiliary matter fields, this modified theory may violate the Strong Equivalence Principle (SEP). It should be noted that in some other alternative theories, SEP can also be violated, e.g., see \cite{2020PrPNP.11203772T} and references therein.  
As the equivalence principle is well constrained by the Solar System experiments \cite{2014LRR....17....4W}, it can impose stringent constraints on these types of theories. 
Also, similar to the other theories, the deviation of this theory from GR should be extremely small in the Solar System scale, otherwise, it cannot survive the classical tests of GR.
Although the deviation is small at this scale, the modified theories can induce large differences at the cosmological scale and be the key to the $\Lambda$CDM enigmas, e.g., see \cite{2004cstg.book.....F}. From this perspective, it is important to constrain the matter-type modified theory and check its validity in the Solar System scale.  

In the current paper, we focus our attention on a specific model of the matter-type modified theories of gravity entitled energy-momentum squared gravity (EMSG). In the EMSG theory, an arbitrary function of the Lorentz scalar ${\boldsymbol{T}}^2=T_{\mu\nu}T ^{\mu\nu}$, i.e., $f({\boldsymbol{T}}^2)$, is added to the EH action. The presence of the square of the energy-momentum tensor in the action and, accordingly, in the field equations fully justifies its appellation.  
Several classes have been introduced for this theory \cite{2014EPJP..129..163K,2016PhRvD..94d4002R,2018PhRvD..97b4011A,2018PhRvD..98f3522A,2019EPJC...79..846A,2020PhRvD.102f4016N} followed by a surge of interest in applying them to various cosmological and astrophysical contexts, for instance,  see  \cite{2018PhRvD..97l4017A,2018PhRvD..98b4031N,2019A&A...625A.127F,2019PhRvD.100h3511B,2020PhRvD.102l4059A,2020PhRvD.101d4058B,2020EPJC...80..150K,2020PhRvD.101f4021C,2021IJMPA..3650004S,2021PDU....3100774S,2021PDU....3300849R,2021EPJP..136..253C,2022PDU....3801128A,2022PDU....3601013K,2022PDU....3701084K,2022AnPhy.44769149T,2022PhRvD.105d4014N,2022PhRvD.105j4026N,2022PDU....3601015Y,2022EPJP..137..322Y,2023EPJC...83.1040L,2023PhRvD.108b3524F,2023MNRAS.523.5452A,2023AnPhy.45869440P,2024PDU....4401463C,2024PDU....4501505A,2024arXiv240201210J}. 
Among the EMSG classes, we investigate a class called the quadratic-EMSG where $f({\boldsymbol{T}}^2)=f_0'{\boldsymbol{T}}^2$ and $f_0'$ is its free parameter\footnote{To coordinate with our previous works, we utilize the same notation for the free parameter of the EMSG theory applied in Ref. \cite{2022PhRvD.105d4014N,2022PhRvD.105j4026N}.}. This class, similar to the other EMSG classes, shows that the effective energy-momentum tensor $T^{\text{eff}}_{\mu\nu}=T_{\mu\nu}+T_{\mu\nu}^{\text{\tiny EMSG}}$ is indeed conserved while violating the standard local conservation law due to nonminimal interaction between $T_{\mu\nu}$ and its EMSG  accompanying partner $T_{\mu\nu}^{\text{\tiny EMSG}}$. In the following,  $T_{\mu\nu}^{\text{\tiny EMSG}}$, which is different in each case, is defined for a perfect fluid in the quadratic-EMSG theory. It should be mentioned that this model can be more effective in the early universe and dense compact astrophysical objects \cite{2016PhRvD..94d4002R}. This class of EMSG theory has been constrained in the strong-gravity regime, e.g., see \cite{2018PhRvD..97l4017A,2018PhRvD..98b4031N,2022PhRvD.105d4014N}.

The Solar System tests are based on the parametrized post-Newtonian (PPN) formalism \cite{will2018theory}, which can be used to test most alternative theories of gravity in the weak-field limit. This formalism is introduced for the point particles \cite{1968PhRv..169.1014N,1968PhRv..169.1017N} and for the perfect-fluid systems \cite{1971ApJ...163..595T,1971ApJ...163..611W,1971ApJ...169..125W,1971ApJ...169..141W}. To analyze this new theory using Solar System experiments, the weak-field limit, more precisely, the first order of the post-Newtonian (PN) limit of the quadratic-EMSG theory should be studied. Then, utilizing the experimental bounds on the PPN parameters obtained in the Solar System scale \cite{2014LRR....17....4W,2003Natur.425..374B,2004PhRvL..92l1101S,2009IJMPD..18.1129W}, the free parameter of the theory can be constrained. In \cite{2022PhRvD.105j4026N}, after introducing the PN expansion of the quadratic-EMSG theory, we derive the PN near-zone metric of a perfect fluid and the null geodesic of the EMSG curved spacetime. Considering a spherically symmetric finite object in this theory, we next study the ``classical" tests including (i) the deflection of light and (ii) the time delay of light. It is shown that the EMSG theory passes
these two Solar System tests with flying colors. Furthermore, it is argued that if the density of the deflector is known, one can distinguish EMSG from GR and restrict its free parameter utilizing the bounds coming from experiments. In \cite{2023PDU....4201305A}, introducing the EMSG spherically symmetric solution in the context of the PPN formalism, local tests are also investigated for the EMSG model of the form $f({\boldsymbol{T}}^2)=\alpha {\boldsymbol{T}}^{2\eta}$ where $\alpha$ and $\eta$ are constants.
It is shown that only if the physical mass of the astrophysical object or the values of $\alpha$ and $\eta$ are known from another window, this theory and GR are distinguishable by Solar System tests.  

Following these studies, in order to obtain the motion of bodies with self-gravitational interactions such as planets and the Sun, and consequently to study the next classical test, the perihelion advance of Mercury, in the framework of the EMSG theory, the $N$-body acceleration needs to be derived. 
It should be noted that since the motion of a test particle and photon in the field of a single spherically symmetric body is studied in \cite{2022PhRvD.105j4026N,2023PDU....4201305A}, their results are not sufficient for this purpose and should be extended.      
One way to obtain the $N$-body acceleration is to treat each body of the system as a ``point particle" and assume that it moves on a test-body geodesic in a curved EMSG spacetime whose PPN metric is produced by all components of the $N$-body system with taking care of the gravitational field of the body itself. But due to the appearance of the quadratic term of density $\rho^2$ in the quadratic-EMSG theory, this description will be quite complicated\footnote{In fact, in this way, one quickly encounters several ill-defined integrals involving the square of the delta function. To overcome this mathematical difficulty, a regularization prescription needs to be introduced. This mathematical issue and related rule for regularizing divergent integrals is not the subject of the present study. So we leave the point-particle model in the EMSG theory for the future.}. On the other hand, the trajectory of the massive self-gravitating body is not necessarily geodesics of any PPN metric and its motion may be influenced by its internal structure (except in GR) \cite{will2018theory}. Besides that, as pointed out earlier, in the matter-type modified theory of gravity, the body's motion depends on the interaction between the standard and modified matter fields and thus does not follow geodesics. Therefore, in general, the trajectory of the body is expected to depend on its internal structure where the EMSG effects may play a role, thereby violating at least one aspect of SEP, namely the Gravitational Weak Equivalence Principle (GWEP). The bottom line is that the $N$-body acceleration may host the EMSG effects which can be constrained in the solar scale using the perihelion shift of the planets as well as experimental tests of SEP/GWEP.

To shed light on these points, we apply the method described in \cite{will2018theory,poisson2014gravity} and present appropriate $N$-body equations of motion in the EMSG theory by considering each body realistically. This method is based on defining the center-of-mass position of the body. It is indeed the main tool for studying the problem of motion \cite{1987thyg.book..128D}. It is assumed that each body is a finite, non-rotating, self-gravitating ball of the PN perfect fluid that is well separated from other bodies and has a weak mutual gravitational interaction. So, for each PN body, the slow-motion condition, $v^2/c^2\ll 1$, and weak-field limit, $U/c^2\ll 1$, are established. Here, $v$ is the velocity of the fluid element and $U$ is the Newtonian potential.  
It is worth noting that the $N$-body problem is also studied in the PN limit of other modified theories of gravity, for instance, see \cite{2022EPJC...82..782B,2023EPJC...83..112B}.
In Sec. \ref{secII}, the exact as well as the PN hydrodynamic equations are obtained in the quadratic-EMSG theory. Armed with the PN EMSG hydrodynamic equations, the equations of motion of each body are derived in Sec. \ref{secIII}. In this section, we introduce the viral identities in this theory and then obtain the corresponding equilibrium conditions. Utilizing these conditions, the final form of the $N$-body acceleration, which is the purpose of this section, is achieved. To complete our results, Sec. \ref{secIV} is dedicated to the derivation of the PN inter-body metric of the $N$-body system in the quadratic-EMSG theory. This metric governs the empty region between the well-separated $N$ bodies/balls up to the first PN order. We discuss our findings in Sec. \ref{secV}. Supplement relations are provided in Appendices \ref{App_PN_limit}-\ref{app3}.

In this paper, $\eta_{\mu\nu}=\text{diag}(-1,1,1,1)$ is the Minkowski metric in Cartesian coordinates, and Greek and Latin indices run over the four spacetime and spatial dimensions, respectively.
We refer to the PN order as $n\text{PN}\equiv O(c^{-2n})$. Also, to simplify PN relations, the relevant Newtonian equations are used.

\section{Energy-momentum squared gravity: field equations}\label{secII}

We launch our study by introducing the field equations of the quadratic-EMSG theory which are derived in \cite{2014EPJP..129..163K,2016PhRvD..94d4002R,2018PhRvD..97b4011A,2017PhRvD..96l3517B}. In preparation for calculations given in the next section, the exact hydrodynamic equations as well as the weak-field hydrodynamic equations will also be derived here.

\subsection{Exact hydrodynamic equations}
The action of the quadratic-EMSG theory is introduced as 
\begin{align}
S=\int \sqrt{-g}\Big(\frac{1}{2k}R+f_0'{\boldsymbol{T}}^2\Big)d^4x+S_{\text{M}},
\end{align}
where $k=\frac{8\pi G}{c^4}$, $g_{\mu\nu}$ is the spacetime metric with the determinant $g$, $R$ is the Ricci scalar, and $S_{\text{M}}$ is the matter action.
Here, $f_0'$ is the free parameter of the theory, which is shown to be positive in \cite{2023PDU....4201360M}, and ${\boldsymbol{T}}^2=T^{\alpha\beta}T_{\alpha\beta}$ where $T_{\alpha\beta}$ is the energy-momentum tensor. 
The quadratic-EMSG field equations are then given by
\begin{align}\label{G}
G_{\mu\nu}=
k\Big( T_{\mu\nu}+f_0'\big( g_{\mu\nu}{\boldsymbol{T}}^2-4T^\sigma_\mu T_{\nu\sigma}-4{\boldsymbol{\Psi}}_{\mu\nu}\big)\Big),
\end{align}
where $G_{\mu\nu}=R_{\mu\nu}-\frac{1}{2}g_{\mu\nu}R$ is the Einstein tensor and   
\begin{align}
{\boldsymbol{\Psi}}_{\mu\nu}=-\mathcal{L}_{\rm m}\big(T_{\mu\nu}-&\frac{1}{2}Tg_{\mu\nu}\big)\\\nonumber
&-\frac{1}{2}TT_{\mu\nu}-2T^{\alpha\beta}\frac{\partial^2 \mathcal{L}_{\rm m}}{\partial g^{\alpha\beta}\partial g^{\mu\nu}},
\end{align}
 in which $\mathcal{L}_{\rm m}$ stands for the matter Lagrangian density and $T$ is the trace of the energy-momentum tensor.
As usual, it is considered that the matter Lagrangian density is only a function of the metric, independent of metric derivatives.

Regarding the field equations \eqref{G} and the contracted Bianchi identities $\nabla_{\mu}G^{\mu\nu}=0$, one can deduce that the following relation
\begin{align}\label{nabla_T_eff}
\nabla_{\mu}T_{\text{eff}}^{\mu\nu}=0,
\end{align} 
is established in this theory. $T_{\text{eff}}^{\mu\nu}$ is the effective conserved energy-momentum tensor, defined as
\begin{align}\label{T_eff}
T^{\text{eff}}_{\mu\nu}=T_{\mu\nu}+T^{\text{\tiny EMSG}}_{\mu\nu},
\end{align}
where
\begin{align}\label{TEMSG}
T^{\text{\tiny EMSG}}_{\mu\nu}=f'_0\Big( g_{\mu\nu}{\boldsymbol{T}}^2-4T^\sigma_\mu T_{\nu\sigma}-4{\boldsymbol{\Psi}}_{\mu\nu}\Big). 
\end{align}
So, unlike GR, the standard energy-momentum tensor is not conserved in EMSG, namely $\nabla_{\mu}T^{\mu\nu}\neq0$. 

We should mention that in addition to the effective energy-momentum conservation \eqref{nabla_T_eff}, we consider that the conservation of the rest-mass density $\rho$ is established in EMSG, i.e., 
\begin{align}
\nabla_{\alpha}\big(\rho u^{\alpha}\big)=0.
\end{align}
Here, $u^{\alpha}=\gamma(c,\boldsymbol{v})$ is the four-velocity field where $\boldsymbol{v}$ is the three-velocity field and $\gamma=u^0/c$. It should be emphasized that the conservation of rest-mass considered here is based on the assumption that the total baryon number of the fluid element is conserved, independent of Eq. \eqref{nabla_T_eff}  and the assumed theory of gravity \cite{will2018theory}. 
This relation can be simplified as  
\begin{align}\label{rho}
\partial_t \rho^*+\partial_j(\rho^* v^j)=0,
\end{align}
where $\rho^*=\sqrt{-g}\gamma \rho$ \cite{poisson2014gravity,1965ApJ...142.1488C}. 
In fact, $\rho^*$ is the rescaled mass density that satisfies the continuity equation \eqref{rho}. In the following calculations, we use this useful definition of the mass density.
Eqs. \eqref{nabla_T_eff} and \eqref{rho} provide us with the exact curved-spacetime hydrodynamic equations in the EMSG theory.

\subsection{Post-Newtonian hydrodynamic equations}\label{PN_hyd}

It is shown that to find the relativistic effects on the motion of a body, more precisely equations of motion for a center of mass of a finite body, in a curved spacetime, we need to know the Euler equation at least up to the first PN (1\tiny PN \normalsize) approximation \footnote{A reader wishing to review the PN limit of the EMSG theory may begin with Sec. III of \cite{2022PhRvD.105j4026N} and then return to this section.}. This point is illustrated in the next section. This order of approximation reveals the very first footprint of the relativistic contributions in the equation of motion.      
We use the results presented in \cite{2022PhRvD.105d4014N,2022PhRvD.105j4026N} to obtain the PN Euler equation in the EMSG theory. In \cite{2022PhRvD.105j4026N}, the near-zone metric of a perfect fluid in EMSG is introduced. For the sake of convenience, the relations required for the present study are summarized in App. \ref{App_PN_limit}. In this appendix, we truncate the components of the metric to the sufficient PN order we need below.  

For the purposes of the current work where the EMSG theory is asked in the Solar System and the stellar system, it is sufficient to assume that the matter part of the system (the matter composing bodies) is a perfect fluid\footnote{In the following sections, it is shown that the final external equations of motion up to the first PN order in the EMSG theory are independent of the internal structure and internal forces of bodies. Therefore, it is sufficient to treat each body as a perfect fluid ball in the case of solar system and stellar system experiments, even though some components of these systems are not exactly fluid.}, whose energy-momentum tensor is defined by
\begin{align}\label{EMT_GR}
T^{\mu\nu}=\Big(\rho+\frac{\epsilon}{c^2}+\frac{p}{c^2}\Big)u^{\mu}u^{\nu}+p\,g^{\mu\nu},
\end{align}
where $\epsilon$ is the proper internal energy density and $p$ is pressure. Inserting this relation into Eq. \eqref{TEMSG}, we get
\begin{align}\label{EMT_EMSG}
\nonumber
&T^{\mu\nu}_{\text{\tiny EMSG}}=f_0'\bigg[\Big(2c^2\rho^2+8\rho\,p+4\rho\,\epsilon+\frac{1}{c^2}\big(6p^2+8p\,\epsilon\\
&+2\epsilon^2\big)\Big)u^\mu u^\nu+\Big(c^4\rho^2+2c^2\rho\,\epsilon+3p^2+\epsilon^2\Big)g^{\mu\nu}\bigg].
\end{align}
To obtain this relation, the normalization condition $g_{\mu\nu}u^{\mu}u^{\nu}=-c^2$ is used. Also it is assumed that the matter Lagrangian density is $\mathcal{L}_{\text{m}}=p$ and the second metric derivative of $\mathcal{L}_{\text{m}}$ vanishes \cite{2017PhRvD..96l3517B,2013PhLB..725..437O,2023arXiv230611717A}.

Now we turn to find the PN Euler equation in the EMSG theory. 
To do so, let us first study the time component of Eq. \eqref{nabla_T_eff}, which can be expanded as 
\begin{align}\label{eq_E}
\nonumber
&\frac{1}{c}\partial_t\big(\sqrt{-g}T^{00}_{\text{eff}}\big)+\partial_j\big(\sqrt{-g}T^{0j}_{\text{eff}}\big)+\Gamma^{0}_{00}\big(\sqrt{-g}T^{00}_{\text{eff}}\big)\\
&+2\Gamma^{0}_{0j}\big(\sqrt{-g}T^{0j}_{\text{eff}}\big)+\Gamma^{0}_{jk}\big(\sqrt{-g}T^{jk}_{\text{eff}}\big)=0,
\end{align}
Inserting Eqs. \eqref{EMT_GR} and  \eqref{EMT_EMSG}, the metric components, and the Christoffel symbols given in App. \ref{App_PN_limit} into the above relation, using Eq. \eqref{rho}, applying the slow-motion condition and weak-field limit, and finally truncating the result to the leading PN order, we find that
\begin{align}\label{energy_eq}
\nonumber
&\rho^*\partial_t\big(\Pi+\frac{1}{2}v^2\big)+\rho^*v^j\partial_j\big(\Pi+\frac{1}{2}v^2\big)+\partial_j\big(p\,v^j\big)\\
&-\rho^*v^j\partial_jU+2c^4f_0'\rho^*v^j\partial_j\rho^*=O(c^{-2}),
\end{align}
where, $\Pi=\epsilon/\rho^*$ is the internal energy of a fluid element which is divided by its mass.
This relation is the local conservation of energy within the fluid in the EMSG theory. To simplify it further, we study the spatial component of Eq. \eqref{nabla_T_eff}, i.e., $\nabla_{\mu}T_{\text{eff}}^{j\mu}=0$. It can be written as follows:
\begin{align}\label{s_com}
\nonumber
&\frac{1}{c}\partial_t\big(\sqrt{-g}T^{0j}_{\text{eff}}\big)+\partial_k\big(\sqrt{-g}T^{jk}_{\text{eff}}\big)+\Gamma^{j}_{00}\big(\sqrt{-g}T^{00}_{\text{eff}}\big)\\
&+2\Gamma^{j}_{0k}\big(\sqrt{-g}T^{0k}_{\text{eff}}\big)+\Gamma^{j}_{kn}\big(\sqrt{-g}T^{kn}_{\text{eff}}\big)=0. 
\end{align}
Keeping only the leading-order terms in the above relation and applying Eq. \eqref{rho}, one can show that the Newtonian limit of the Euler equation in the EMSG theory is given by
\begin{align}\label{Euler-N-EMSG}
\rho^*\frac{dv^j}{dt}=\rho^*\partial_jU-\partial_jp-2f_0'c^4\rho^*\partial_j\rho^*+O(c^{-2}).
\end{align}
Regarding the point mentioned in App. \ref{App_PN_limit}, the extra EMSG term appearing in the Euler equation is of the same order as the standard ones, i.e., it is of the order $c^0$. 
Here, the definition of the total time derivative $d/dt=\partial_t+v^k\partial_k$ is utilized.  
Now substituting Eq. \eqref{Euler-N-EMSG} within Eq. \eqref{energy_eq} and again utilizing Eq. \eqref{rho}, we find the first law of thermodynamics for the isentropic fluid
\begin{align}\label{first-law}
\frac{d\Pi}{dt}=\frac{p}{{\rho^*}^2}\frac{d\rho^*}{dt}+O(c^{-2}),
\end{align}
in this theory. As seen, in this order, the EMSG corrections play no role. 

To obtain the PN Euler equation, we Follow the method applied in \cite{poisson2014gravity,will2018theory} in the PN framework. By inserting the results given in App. \ref{App_PN_limit} as well as Eqs. \eqref{EMT_GR}, \eqref{EMT_EMSG}, and \eqref{first-law} into Eq. \eqref{s_com}, after some manipulations, we finally arrive at 
\begin{align}\label{Euler-PN-EMSG}
\nonumber
&\rho^*\frac{dv^j}{dt}=-\partial_jp+\rho^*\partial_jU+\frac{1}{c^2}\bigg\lbrace\Big(\Pi+\frac{p}{\rho^*}+\frac{1}{2}v^2+U\Big)\partial_jp\\\nonumber
&-v^j\partial_tp+\rho^*\Big[\big(v^2-4U\big)\partial_jU-v^j\big(3\partial_tU+4v^k\partial_kU\big)\\\nonumber
&+4v^k\big(\partial_kU_j-\partial_jU_k\big)+4\partial_tU_j+\partial_j\Psi\Big]\bigg\rbrace-2f_0'c^4\rho^*\bigg\lbrace\partial_j\rho^*\\\nonumber
&-\frac{1}{c^2}\Big[v^j\big(v^k\partial_k\rho^*+\rho^*\partial_kv^k\big)-\big(\Pi-\frac{3}{2}v^2-7U\big)\partial_j\rho^*\\\nonumber
&+\frac{1}{\rho^*}\partial_j(\rho^*p)+2f_0'c^4\rho^*\partial_j\rho^*-\rho^*\partial_j\big(\Pi-\frac{1}{2}v^2-3U\big)\\
&+2\partial_jU_{\text{\tiny EMSG}}\Big]\bigg\rbrace+O(c^{-4}).
\end{align}  
Here, the definition of the total time derivative is utilized and the terms are kept to $O(c^{-2})$. This is the PN version of the Euler equation in the EMSG theory. It is seen that in addition to the PN corrections related to GR, i.e., the terms within the first braces, those related to EMSG, i.e., the terms within the second braces, appear in this equation. 
It should be noted that in the terms $\partial_tU_j$ and $\partial_j\Psi$, other EMSG effects are also hidden. In the following section, these hidden effects are derived and their role is examined in the final result.

\section{Dynamics of a body in energy-momentum squared gravity}\label{secIII}

Utilizing the PN Euler equation \eqref{Euler-PN-EMSG}, we now study the external dynamics of a body with the relativistic effects to the 1\tiny PN \normalsize order in the EMSG theory. 
To do so, we consider the coarse-grained description, e.g., see \cite{will2018theory,poisson2014gravity}. In this description, a fluid distribution breaks up into a collection of well-separated $N$ bodies moving under the influence of their mutual interactions. The region between bodies is empty and there is no flux of matter from the bodies. Each body is assumed to be 
a finite, non-rotating, self-gravitating ball of a perfect fluid with zero pressure on its surface whose internal dynamics are governed by the PN EMSG hydrodynamic equations.  
It should be noted that the internal dynamics of all bodies are assumed to be governed by the same Eq. \eqref{Euler-PN-EMSG} with the same EMSG free parameter. That is, the EMSG field is the same for all bodies. This is a suitable assumption to avoid adding $N$ different EMSG free parameters to the motion problem. Moreover, as we consider a self-gravitating body, each gravitational potential in the Euler equation \eqref{Euler-PN-EMSG} is produced partly by the body itself, labeled by $A$, and partly by the rest of the bodies in the $N$-body system. So, one can decompose the total gravitational potential, let us call it $\phi$ generally, as $\phi=\phi_A+\phi_{-A}$ where $\phi_A$ and $\phi_{-A}$ are the internal potential, due to the body $A$, and the external potential, due to the rest of the bodies, respectively \footnote{ Interested readers can see the potential decomposition technique in detail in \cite{poisson2014gravity}. The general form of $\phi_A$ and $\phi_{-A}$ is given in App. \ref{app3}.}.

More importantly, as mentioned, it is assumed that the bodies are well-separated. 
That is, the inequality  $R\ll r$ is established. Here, $R$ is the typical size of the body and $r$ is the typical separation distance between bodies making up the $N$-body system. In the following calculation, all terms of the order $\big(R/r\big)^2$ are then discarded. So, the finite-size effects of bodies, such as the multipole moment terms, are not important in this study.  
This important assumption results that the internal and external dynamics respectively proceeding on the time scales $T_{\text{int}}\sim(R^3/G\,m)^{1/2}$ and $T_{\text{ext}}\sim(r^3/G\,m)^{1/2}$ take place over widely separated time scales, i.e., $T_{\text{int}}\ll T_{\text{ext}}$. Here, $m$ is the mass of the body. In fact, the strong inequality $T_{\text{int}}\ll T_{\text{ext}}$ is established for the well-separated bodies with $R\ll r$  \cite{poisson2014gravity,1987thyg.book..128D}. This leads us to the conclusion that the external and internal dynamics are well decoupled and each body can be balanced under its own internal dynamics, that is, it is in dynamical equilibrium.

In this description, center-of-mass variables for each body are involved in the problem of motion. So, for clarity, let us recall here these definitions to the leading PN order. We will introduce their generalized form in the EMSG theory in the following subsections. In this framework, the material mass of a body, denoted by the index $A$, is given by \footnote{In the post-Newtonian gravity, instead of the usual mass density $\rho$, the rescaled mass density defined by $\rho^*=\sqrt{-g}\gamma \rho$ is utilized. It is due to the form of the equation of mass conservation \eqref{rho} in this framework \cite{1965ApJ...142.1488C}. In the Newtonian limit, it is simply replaced by $\rho$.}
\begin{align}\label{m_A}
 m_A=\int_{A}\rho^*d^3x.
\end{align}
The integration domain $V_A$ is assumed to be a time-independent region extending beyond the volume occupied by the body so that in a time interval $dt$, the body does not cross the border of $V_A$. At the same time, the volume $V_A$ is considered to be small enough to neither contain nor interact with other bodies of the system. One can then define the position of the center-of-mass of body $A$ as
\begin{align}\label{r_A}
r_A^j=\frac{1}{m_A}\int_A\rho^* x^jd^3x,
\end{align}
in the global inertial frame. 
Using this definition, we can then proceed to derive the center-of-mass velocity
\begin{align}\label{v_A}
\nonumber
v_A^j&=\frac{d r^j_A}{dt}\\
&=\frac{1}{m_A}\int_A\rho^*v^jd^3x,
\end{align}
considered as the velocity of the body, and the center-of-mass acceleration
\begin{align}\label{a_j-EMSG}
\nonumber
a_A^j&=\frac{d v^j_A}{dt}\\
&=\frac{1}{m_A }\int_A\rho^*\frac{dv^j}{dt}d^3x,
\end{align}
as the body's acceleration. To obtain these relations, we use Eq. \eqref{rho} and the fact that $dm_A/dt=0$. The next step to finding the equations of motion is to insert the Euler equation into the above relation and evaluate each term. For the sake of simplicity, it is considered that each body is reflection-symmetric about its center-of-mass \footnote{Considering the high degree of symmetry of the bodies in the Solar System, this assumption is adequate for our purpose in this paper.}. This is a reasonable assumption that significantly reduces the amount of computation.

For the following calculations, let us also introduce here two useful relative vectors
\begin{align}
\label{rela-quan}
\bar{\boldsymbol{x}}\equiv\boldsymbol{x}-\boldsymbol{r}_{A}(t),~~~~~~\bar{\boldsymbol{v}}\equiv\boldsymbol{v}-\boldsymbol{v}_{A}(t),
\end{align}
which are the position of a fluid element relative to the center-of-mass of the body $\boldsymbol{r}_A(t)$, and the velocity of the element relative to the body velocity $\boldsymbol{v}_{A}(t)$, respectively.  Based on the relative vector $\bar{\boldsymbol{x}}$, the mathematical form of the reflection-symmetric assumption is 
\begin{align}\label{refl}
y (t,\boldsymbol{r}_A-\bar{\boldsymbol{x}})=y(t,\boldsymbol{r}_A+\bar{\boldsymbol{x}}),
\end{align}  
where $y$ represents one of the fluid variables $\rho^*$, $p$, and $\Pi$.  

\subsection{Crude estimation}\label{crude}
As a warm-up exercise, let us first describe what is required to obtain the dynamics of a body that is a member of an $N$-body system and estimate the role of the EMSG terms.
To do so, we use the definition of the body's acceleration obtained in GR, i.e., Eq. \eqref{a_j-EMSG}. Regarding this relation, we know that the weighted average of the accelerations of individual fluid elements inside a body given with respect to an origin outside the body, i.e., $d\boldsymbol{v}/dt$, provides us with the acceleration of the body as whole with respect to that origin outside.
On the other hand, from Eq. \eqref{Euler-PN-EMSG}, it is deduced that several EMSG terms affect the fluid element acceleration  $dv^j/dt$. Due to the complexity of this relation, extracting the full EMSG effects is a complicated task. We derive it in detail in the following section. For now, in this exercise, we consider the Newtonian Euler equation together with only one relativistic part of the EMSG expressions, namely, $4f_0'c^2\rho^*\partial_jU_{\text{\tiny EMSG}}$. We then have 
\begin{align}
\rho^*\frac{dv^j}{dt}\approx\rho^*\partial_jU-\partial_jp-2f_0'c^4\rho^*\Big(\partial_j\rho^*-\frac{2}{c^2}\partial_jU_{\text{\tiny EMSG}}\Big).
\end{align}
As an estimate, inserting the above relation into the definition of $\boldsymbol{a}_A$ and then decomposing the gravitational potentials, after some simplifications, we get
\begin{align}\label{a-estimate}
&\int_A\rho^*\big(dv^j/dt\big)d^3x=\int_A\rho^*\Big[\partial_jU_A+\partial_jU_{-A}\\\nonumber
&~~~~~~~~~~+4f_0'c^2\big(\partial_jU_{\text{\tiny EMSG}, A}+\partial_jU_{\text{\tiny EMSG}, -A}\big)\Big]d^3x,
\end{align} 
where 
\begin{subequations}
\begin{align}
\label{UA}
& U_A=G\int_{A}\frac{{\rho^*}'}{\rvert{\boldsymbol{x}-\boldsymbol{x}'}\rvert}d^3x',\\
& U_{-A}=\sum_{B\neq A}G\int_{B}\frac{{\rho^*}'}{\rvert{\boldsymbol{x}-\boldsymbol{x}'}\rvert}d^3x',
\end{align}
\end{subequations}
are the internal and external pieces of the standard potential $U$ of the $N$-body system, respectively, and   
\begin{subequations}
\begin{align}
& U_{\text{\tiny EMSG},A}=G\int_{A}\frac{{{\rho^*}'}^2}{\rvert\boldsymbol{x}-\boldsymbol{x}'\rvert}d^3x',\\
\label{U-AEMSG}
& U_{\text{\tiny EMSG},-A}=\sum_{B\neq A}G\int_{B}\frac{{{\rho^*}'}^2}{\rvert\boldsymbol{x}-\boldsymbol{x}'\rvert}d^3x',
\end{align}
\end{subequations}
are the internal and external pieces of $U_{\text{\tiny EMSG}}$, respectively. Here, we use the fact that the pressure on the surface of the body is zero. Substituting the internal piece of $U$ and $U_{\text{\tiny EMSG}}$ in Eq. \eqref{a-estimate}, one can show that these parts also vanish. Therefore, the equations of motion reduces to 
\begin{align}
\boldsymbol{a}_A\propto\int_A\rho^*\Big[\partial_jU_{-A}
+4f_0'c^2\partial_jU_{\text{\tiny EMSG}, -A}\Big]d^3x.
\end{align}
This estimate reveals the possible footprint of the EMSG term in the acceleration of body $A$. 
So, the external equations of motion of the $N$-body system may host the EMSG effects.   
To have a proper analysis, we should however first define a suitable center-of-mass position and consequently a center-of-mass acceleration in the EMSG theory, and then consider the rest of the EMSG terms in the fluid dynamics to obtain the complete external equations of motion in the EMSG framework.

\subsection{Equations of motion of the $N$-body system}

To obtain an appropriate definition for the center-of-mass variables in the PN limit of EMSG, we follow the approach utilized in \cite{1974exgr.conf....1W,1976ApJ...205..592C,1977GReGr...8..463S,will2018theory} and generalize Eqs. \eqref{m_A} and \eqref{r_A} in the EMSG framework.
We first define the total mass-energy of body $A$ as $M_A\equiv m_A+E_A/c^2$ where  
$E_A$ is the total energy of body which should be obtained from Eq. \eqref{energy_eq} by taking into account the possible role of EMSG terms. We integrate this equation over the volume occupied by body $A$. After using Eq. \eqref{rho} and the definition of the total time derivative, we find that to the first PN accuracy,  
\begin{align}\label{dE_A}
\frac{dE_A}{dt}=0,
\end{align}
where 
\begin{align}\label{total_energy}
E_A=T_A+\Omega_A+E^{\text{int}}_A+f_0'c^4\mathfrak{M}_A+O(c^{-2}).
\end{align}
The definition of the kinetic energy $T_A$, the gravitational potential energy $\Omega_A$, the internal energy $E^{\text{int}}_A$, and the EMSG portion to energy $\mathfrak{M}_A$ are given in App. \eqref{app2}. From the physical point of view, the last term in Eq. \eqref{total_energy} can be interpreted as the potential energy of the body that is related to the interaction between the standard and modified matter fields in the EMSG theory. 
On the other hand, making use of Eq. \eqref{rho} again, it can be shown that $m_A$ is also constant with respect to time, $dm_A/dt=0$, as long as there is no matter flux from the body $A$.  
Regarding this fact as well as considering Eq. \eqref{dE_A}, it is deduced that $dM_A/dt=0$. 
Gathering the integral definitions \eqref{m_A}, \eqref{T_A}, \eqref{Omega_A}, \eqref{mEMSG}, and \eqref{E_A} together, the integral form of the total mass-energy of body $A$ is written as  
\begin{align}
\nonumber
&M_A\equiv\int_A\rho^*\Big[1+\frac{1}{c^2}\big(\Pi+\frac{1}{2}\bar{v}^2-\frac{1}{2}U_A\\
&~~~~~~~~~~~~~~~~~~~~+f_0'c^4\rho^*\big)\Big]d^3x+O(c^{-4}),
\end{align}
by which one then defines the center-of-mass position as
\begin{align}
\nonumber
&\boldsymbol{R}_A\equiv\frac{1}{M_A}\int_A\rho^*\boldsymbol{x}\Big[1+\frac{1}{c^2}\big(\Pi+\frac{1}{2}\bar{v}^2-\frac{1}{2}U_A\\
&~~~~~~~~~~~~~~~~~~~~~~~~~+f_0'c^4\rho^*\big)\Big]d^3x+O(c^{-4}).
\end{align}

Utilizing this definition, we can now obtain the acceleration of the center-of-mass of body $A$ as $\boldsymbol{a}_A=d^2\boldsymbol{R}_A/dt^2$ in the EMSG theory. Before doing so, let us recall the assumption of the reflection symmetry mentioned earlier. Imposing Eq. \eqref{refl} shows that $\boldsymbol{R}_A=\boldsymbol{r}_A+O(c^{-4})$.
Since our aim is to study the motion of the body to the first PN order, hereafter, we restrict ourselves to the definition \eqref{r_A} providing sufficient information with the desired accuracy. Knowing that $dm_A/dt=0$, we then recover Eqs. \eqref{v_A} and \eqref{a_j-EMSG} for the center-of-mass velocity and acceleration to the first PN limit of the EMSG theory. 
Thanks to the reflection symmetry applied here, the final result for $\boldsymbol{a}_A$ is not too complicated. It should be noted general systems that are not necessarily reflection-symmetric are investigated in \cite{Nazari2023EMSG} to examine other possible EMSG effects on the acceleration of bodies.   

The remaining job toward deriving the coordinate acceleration of the center-of-mass of body $A$ is to insert the PN Euler equation \eqref{Euler-PN-EMSG} within Eq. \eqref{a_j-EMSG} and simplify the result.  By doing so, we arrive at  
\begin{align}
m_A a_A^j=F_0^j+\sum_{n=1}^{30}F^j_n+O(c^{-4}),
\end{align}
where the force integrals $F_0^j\textendash F_{30}^j$ are given in Eqs. \eqref{F0}-\eqref{F18} and \eqref{F19}-\eqref{F30} in App. \ref{app3}. In this derivation, we benefit from the auxiliary potentials $\Phi_1\textendash \Phi_6$ given in Eqs. \eqref{Phi_1}-\eqref{Phi_6} and rewrite $\Psi$ in terms of them as $\Psi=\frac{3}{2}\Phi_1-\Phi_2+\Phi_3+3\Phi_4+\frac{1}{2}\partial_{tt}X$ where $\partial_{tt}X= \Phi_1+2\Phi_4-\Phi_5-\Phi_6+2f_0'c^4U_{\text{\tiny EMSG}}$. It should be noted that to obtain $\partial_{tt}X$, Eq. \eqref{Euler-N-EMSG} is used. One of the hidden EMSG effects is then extracted, manifesting itself in $\Psi$ as claimed before. These force integrals are categorized into two groups: the GR class ($F_0^j\textendash F_{18}^j$) and the EMSG class ($F_{19}^j\textendash F_{30}^j$). We should mention that there is still another hidden EMSG effect in the integral force $F_{12}^j$ which will be taken into account in the following calculations.

We next simplify each force integral by applying the assumptions mentioned at the beginning of this section and also dividing the Newtonian, PN, and EMSG gravitational potentials into two parts: the portion produced by body $A$, which is called the internal potential, and those produced by the remaining bodies in the $N$-body system, which are called the external potentials, e.g., cf. Eqs. \eqref{UA}-\eqref{U-AEMSG}. The final results are summarized in Eqs. \eqref{F0f}-\eqref{F30f}. Collecting these forces, we finally obtain
\begin{widetext}
\begin{align}\label{a_A}
\nonumber
&a_A^j=\partial_jU_{-A}+\frac{1}{m_A\,c^2}\bigg\lbrace v^k_A\Big[2L_A^{(jk)}+3K_A^{jk}-4H_A^{(jk)}-\delta^{jk}\frac{d}{dt}{P}_A+4f_0'c^4Q_A^{(jk)}-2f_0'c^4\delta^{jk}Q_A\Big]+\partial_jU_{-A}\Big[3P_A+2\mathcal{T}_A\\\nonumber
&+\Omega_A+3f_0'c^4\mathfrak{M}_A\Big]-4\partial_kU_{-A}\Big[\Omega^{jk}_A+2\mathcal{T}^{jk}_A+\delta^{jk}P_A+f_0'c^4\delta^{jk}\mathfrak{M}_A\Big]\bigg\rbrace+\frac{1}{c^2}\bigg\lbrace\partial_jU_{-A}\Big(v_A^2-4U_{-A}\Big)-v_A^j\Big(3\partial_tU_{-A}\\\nonumber
&+4v_A^k\partial_kU_{-A}\Big)-4v_A^k\Big(\partial_jU_{k,-A}-\partial_kU_{j,-A}\Big)+4\partial_tU_{j,-A}+2\partial_j\Phi_{1,-A}-\partial_j\Phi_{2,-A}+\partial_j\Phi_{3,-A}+4\partial_j\Phi_{4,-A}\\
&-\frac{1}{2}\partial_j\Phi_{5,-A}-\frac{1}{2}\partial_j\Phi_{6,-A}+5f_0'c^4\partial_jU_{\text{\tiny EMSG},-A}\bigg\rbrace+O(c^{-4}).
\end{align}
\end{widetext}
Here, $A^{(j}B^{k)}=\frac{1}{2}\big(A^jB^k+A^kB^j\big)$. The definitions of the scalar and tensorial quantities appeared in the above relation are given in App. \ref{app2}. 
As seen, several EMSG corrections appear in the acceleration of body $A$. The terms written in the first braces of the above equation depend on the body's internal structure, while the first term $\partial_jU_{-A}$ along with the terms in the second braces are all related to the gravitational potentials of the other bodies in the system.  As grasped from the terms in the first braces, several EMSG corrections may contribute to the violation of GWEP.  

It should be noted that the assumption of dynamical equilibrium of body $A$ has not been imposed up to this point, which can in principle eliminate some of these effects. 
In fact, this assumption can help us to further simplify the above complicated result. 
Since the fluid dynamics in the EMSG theory changes even in the Newtonian order as shown in Eq. \eqref{Euler-N-EMSG}, we should first study the viral identities in this theory and then obtain the corresponding equilibrium conditions. In general, these conditions can be affected by the EMSG corrections. This point is the subject of the following section. 

\subsection{Virial identities} 
To introduce the virial integrals in the EMSG theory, we start from the first time derivative of the quadrupole-moment tensor $I^{jk}$ defined in Eq. \eqref{I^jk}. Considering the non-spinning body, for which the spin tensor defined in Eq. \eqref{S^jk} vanishes $S^{jk}_A=0$, one can show that
\begin{align}
\label{dotI}
\frac{1}{2}\frac{d}{dt}{I}^{jk}_A= \int_A\rho^*\bar{x}^k\bar{v}^j\,d^3x.
\end{align}  
The next virial identity in the framework EMSG theory is obtained from the second time derivative of $I^{jk}$ as
\begin{align}\label{ddotI}
\frac{1}{2}\frac{d^2}{dt^2}{I}^{jk}_A=2\mathcal{T}^{jk}_A+\Omega^{jk}_A+\delta^{jk}P_A+f_0'c^4\delta^{jk}\mathfrak{M}_A+O(c^{-2}).
\end{align}
To derive this relation, Eqs. \eqref{Euler-N-EMSG} and \eqref{r_A} are used. Moreover, we utilize the definition of the acceleration of the fluid element relative to the body acceleration, $\bar{a}^j=d\bar{v}^j/dt=dv^j/dt-a_A^j$, obtained from Eq. \eqref{rela-quan}. And finally by taking three derivatives of ${I}^{jk}_A$, we deduce that 
\begin{align}\label{dddotI}
&\frac{1}{2}\frac{d^3}{dt^3}{I}^{jk}_A=4H^{(jk)}_A-2L_A^{(jk)}+\delta^{jk}\frac{d}{dt}{P}_A-3K_A^{(jk)}\\\nonumber
&~~~~~~~~~~~-f_0'c^4\big(4Q^{(jk)}_A-2\delta^{jk}Q_A\big)+O(c^{-2}),
\end{align}
where the relation $d\mathcal{T}^{jk}_A/dt=H_A^{(jk)}-L_A^{(jk)}-2f_0'c^4Q_A^{(jk)}$ as well as Eq. \eqref{v_A} are used.      
Notice that due to the required accuracy in the current study, these identities are derived up to the leading (Newtonian) order, $O(1)$. Furthermore, based on the assumption that the $N$-body system consists of  well-separated bodies, the terms related to the external gravitational potentials which are of the order $(R/r)^2$ compared to the leading terms, are reasonably removed from Eqs. \eqref{ddotI} and \eqref{dddotI}.

We are now in a position to derive the equilibrium conditions in the EMSG theory. To do so, we  turn to the dynamical equilibrium assumption applied to each body of the $N$-body system. This assumption in fact means that the structure properties of bodies are time-independent so that the total time derivatives of any internal quantity like the quadrupole-moment tensor vanishes. Consequently, all terms on the left-hand side of Eqs. \eqref{dotI}-\eqref{dddotI} become zero.
Therefore, the equilibrium conditions in the EMSG theory will be
\begin{subequations}
\begin{align}
\label{second-eq-condi}
&0=4H^{(jk)}_A-2L_A^{(jk)}+\delta^{jk}\frac{d}{dt}{P}_A-3K_A^{(jk)}\\\nonumber
&~~~~~~~~~~~~~~~~-f_0'c^4\big(4Q^{(jk)}_A-2\delta^{jk}Q_A\big)+O(c^{-2}),\\
\label{first-eq-condi}
&0=2\mathcal{T}^{jk}_A+\Omega^{jk}_A+\delta^{jk}P_A+f_0'c^4\delta^{jk}\mathfrak{M}_A+O(c^{-2}).
\end{align}
\end{subequations}
For the following calculation, the trace of Eq. \eqref{first-eq-condi} is also obtained below   
\begin{align}
\label{third-eq-condi}
0=2\mathcal{T}_A+\Omega_A+3\,P_A+3\,f_0'c^4\mathfrak{M}_A +O(c^{-2}).
\end{align}
These results show that the EMSG corrections manipulate the equilibrium conditions.
As the final point of this part, we should mention that similar to the GR case, the first virial identity \eqref{dotI} reveals that 
\begin{align}\label{condi1}
 \int_A\rho^*\bar{x}^k\bar{v}^j\,d^3x=0,
\end{align}
for a non-spinning body in dynamical equilibrium. 

\subsection{Final form of the acceleration}

Armed with the equilibrium conditions \eqref{second-eq-condi}, \eqref{first-eq-condi}, and \eqref{third-eq-condi}, we simplify the acceleration \eqref{a_A} as
\begin{align}\label{a_n-body}
&a^j_A=\partial_jU_{-A}+\frac{1}{c^2}\bigg\lbrace\partial_jU_{-A}\Big(v_A^2-4U_{-A}\Big)\\\nonumber
&-v_A^j\Big(3\partial_tU_{-A}+4v_A^k\partial_kU_{-A}\Big)-4v_A^k\Big(\partial_jU_{k,-A}-\partial_kU_{j,-A}\Big)\\\nonumber
&+4\partial_tU_{j,-A}+2\partial_j\Phi_{1,-A}-\partial_j\Phi_{2,-A}+\partial_j\Phi_{3,-A}+4\partial_j\Phi_{4,-A}\\\nonumber
&-\frac{1}{2}\partial_j\Phi_{5,-A}-\frac{1}{2}\partial_j\Phi_{6,-A}+5f_0'c^4\partial_jU_{\text{\tiny EMSG},-A}\bigg\rbrace+O(c^{-4}).
\end{align}
This is the equation of motion of body $A$ among a system of $N$ gravitating bodies which are in dynamical equilibrium. Here, all external potentials are evaluated at $\boldsymbol{x}=\boldsymbol{r}_A (t)$ after differentiation. As seen, the motion of body $A$ depends not only on the external piece of the gravitational potentials given in GR (i.e., $U_{-A}$, $U_{-A}^j$, $\Phi_{n,-A}$ where $n=1,\dots,6$) but also on the external piece of the EMSG gravitational potential, $U_{\text{\tiny EMSG},-A}$. This point has also been predicted in the previous section \ref{crude} using a crude estimate. 
It is obvious that the terms $O(c^{-2})$ in the equilibrium conditions eventually contribute to $O(c^{-4})$ in the above relation. This is why we do not take them into account in Eqs. \eqref{second-eq-condi}, \eqref{first-eq-condi}, and \eqref{third-eq-condi}. 
Furthermore, imposing the equilibrium conditions results in the acceleration in which the internal structure of the body, especially those related to EMSG corrections, no longer plays a role.

As the last step toward finding the final form of $\boldsymbol{a}_A$, 
the external piece of all gravitational potentials in Eq. \eqref{a_n-body} should be obtained. We follow the method introduced in chapter 9 of \cite{poisson2014gravity} to do so\footnote{Since the method for deriving the external potentials is described in detail in \cite{poisson2014gravity}, we refrain from repeating it here and refer the interested reader to this reference.}.    
After some algebra, we come to the conclusion that the EMSG effects play a role only in $\partial_tU^j_{-A}$ as well as $\partial_jU_{\text{\tiny EMSG},-A}$. These terms are respectively obtained as
\begin{align}\label{UjwithEMSG}
\nonumber
&\partial_tU^j_{-A}=\sum_{B\neq A}G\Big(2\mathcal{T}^{jk}_B+\Omega^{jk}_B+\delta^{jk}P_B+f_0'c^4\delta^{jk}\mathfrak{M}_B\Big)\frac{n^k_{AB}}{r^2_{AB}}\\\nonumber
&+\sum_{B\neq A}\frac{Gm_B\big(\bm{n}_{AB}\cdot\bm{v}_B\big)v^j_B}{r^2_{AB}}+\sum_{B\neq A}\frac{G^2m_Am_Bn^j_{AB}}{r^3_{AB}}\\
&-\sum_{B\neq A}\sum_{C\neq A,B}\frac{G^2 m_Bm_Cn^j_{BC}}{r_{AB}r^2_{BC}},
\end{align}
and 
\begin{align}\label{U_EMSG_ext}
\partial_jU_{\text{\tiny EMSG},-A}=-\sum_{B\neq A}\frac{G\,\mathfrak{M}_Bn^j_{AB}}{r^2_{AB}},
\end{align}
where $\bm{r}_{AB}\equiv \bm{r}_A-\bm{r}_B$, $r_{AB}\equiv|\bm{r}_{AB}|$, and $\bm{n}_{AB}\equiv\frac{\bm{r}_{AB}}{r_{AB}}$. Eq. \eqref{UjwithEMSG} in fact indicates the last hidden EMSG effect mentioned previously, which is indeed present in the integral force $F^j_{12}$. So, up to this point, all EMSG contributions have been considered. 
The rest of the external gravitational potentials are similar to those obtained in GR. We summarize the results below
\begin{subequations}
\begin{align}
\label{par_jU_A}
&\partial_jU_{-A}=-\sum_{B\neq A}\frac{G\,m_Bn^j_{AB}}{r_{AB}^2},\\
&\partial_tU_{-A}=\sum_{B\neq A}\frac{G\,m_B(\boldsymbol{n}_{AB}\cdot\boldsymbol{v}_B)}{r_{AB}^2},\\
&\partial_kU^j_{-A}=-\sum_{B\neq A}\frac{G\,m_B\,v^j_B\,n_{AB}^k}{r^2_{AB}},\\
&\partial_j\Phi_{1,-A}=-\sum_{B\neq A}\frac{2\,G\,\mathcal{T}_B\,n^j_{AB}}{r_{AB}^2}-\sum_{B\neq A}\frac{G\,m_B\,v_B^2n^j_{AB}}{r_{AB}^2},\\\nonumber
&\partial_j\Phi_{2,-A}=\sum_{B\neq A}\frac{2\,G\,\Omega_B\,n^j_{AB}}{r_{AB}^2}-\sum_{B\neq A}\frac{G^2\,m_A\,m_B\,n^j_{AB}}{r_{AB}^3}\\
&~~~~~~~~~~~~-\sum_{B\neq A}\sum_{C\neq A,B}\frac{G^2\,m_B\,m_C\,n^j_{AB}}{r^2_{AB}r_{BC}},\\
&\partial_j\Phi_{3,-A}=-\sum_{B\neq A}\frac{G\,E^{\text{int}}_B\,n^j_{AB}}{r^2_{AB}},\\
&\partial_j\Phi_{4,-A}=-\sum_{B\neq A}\frac{G\,P_B\,n^j_{AB}}{r_{AB}^2},\\
&\partial_j\Phi_{5,-A}=-\sum_{B\neq A}G\,\Omega_B^{kn}\partial_{jkn}r_{AB}\\\nonumber
&-\sum_{B\neq A}\sum_{C\neq A,B}\frac{G^2\,m_B\,m_C}{r_{AB}r^2_{BC}}\Big[n^j_{BC}-\big(\boldsymbol{n}_{AB}\cdot\boldsymbol{n}_{BC}\big)n^j_{AB}\Big],\\\nonumber
&\partial_j\Phi_{6,-A}=-\sum_{B\neq A}2\,G\mathcal{T}_{B}^{kn}\partial_{jkn}r_{AB}-\sum_{B\neq A}\frac{2\,G\,\mathcal{T}_Bn^j_{AB}}{r^2_{AB}}\\
\label{parjPhi6_A}
&+\sum_{B\neq A}\frac{G\,m_B\big(\boldsymbol{n}_{AB}\cdot\boldsymbol{v}_B\big)}{r_{AB}^2}\Big[2\,v^j_B-3\big(\boldsymbol{n}_{AB}\cdot\boldsymbol{v}_B\big)n^j_{AB}\Big].
\end{align}
\end{subequations}

Inserting Eqs. \eqref{UjwithEMSG}, \eqref{U_EMSG_ext}, and \eqref{par_jU_A}-\eqref{parjPhi6_A} within Eq. \eqref{a_n-body}, after some manipulations, we arrive at
\begin{align}\label{aA}
\nonumber
\bm{a}_A=&\bm{a}_A[0\text{\tiny PN\normalsize}]+\bm{a}_A^{\rm GR}[1\text{\tiny PN\normalsize}]+\bm{a}_A^{\rm EMSG}[1\text{\tiny PN\normalsize}]\\
&+\bm{a}_A[\text{STR}]+O(c^{-4}),
\end{align}
where
\begin{align}\label{a-N}
a^j_A[0\text{\tiny PN\normalsize}]=-\sum_{B\neq A}\frac{G\,m_Bn^j_{AB}}{r^2_{AB}},
\end{align}
is the Newtonian acceleration, 
\begin{align}\label{aPNGR}
\nonumber
&a^{\text{GR},j}_A[1\text{\tiny PN\normalsize}]=-\frac{1}{c^2}\bigg\lbrace\sum_{B\neq A}\frac{G\,m_B}{r^2_{AB}}\Big[v_A^2-4\big(\bm{v}_A\cdot\bm{v}_B\big)+2v_B^2\\\nonumber
&-\frac{3}{2}\big(\bm{n}_{AB}\cdot\bm{v}_B\big)^2-\frac{5\,G\,m_A}{r_{AB}}-\frac{4\,G\,m_B}{r_{AB}}\Big]n^j_{AB}\\
&-\sum_{B\neq A} \frac{G\,m_B}{r^2_{AB}}\Big[\bm{n}_{AB}\cdot\big(4\bm{v}_A-3\bm{v}_B\big)\Big]\big(v^j_A-v^j_B\big)\\\nonumber
&-\sum_{B\neq A}\sum_{C\neq A,B}\frac{G^2 m_B\,m_C}{r^2_{AB}}\Big[\frac{4}{r_{AC}}+\frac{1}{r_{BC}}-\frac{r_{AB}}{2r^2_{BC}}\\\nonumber
&\times\big(\bm{n}_{AB}\cdot\bm{n}_{BC}\big)\Big]n^j_{AB}+\frac{7}{2}\sum_{B\neq A}\sum_{C\neq A,B}\frac{G^2m_Bm_C}{r_{AB}r^2_{BC}}n^j_{BC}\bigg\rbrace,
\end{align}
is the GR portion of the PN acceleration, and  
\begin{align}\label{a-EMSG}
a_A^{\text{EMSG},j}[1\text{\tiny PN\normalsize}]=-5f_0'c^2\sum_{B\neq A}\frac{G\,\mathfrak{M}_Bn^j_{AB}}{r^2_{AB}},
\end{align}
is the EMSG portion of the PN acceleration. We recall that the relativistic corrections are preserved up to the first PN order, $O(c^{-2})$. Moreover, the structure portion $a^j_A[\text{STR}]$ is obtained as 
\begin{align}\label{a-STR}
&a^j_A[\text{STR}]=-\frac{1}{c^2}\sum_{B\neq A}G\Big[E_B-5f_0'c^4\mathfrak
{M}_B\Big]\frac{n^j_{AB}}{r^2_{AB}},
\end{align}
in which $E_{B}$ and $\mathfrak{M}_B$ are purely determined by the internal structure of body $B$. To extract the above fragment of the acceleration, we again apply the equilibrium conditions \eqref{first-eq-condi} and \eqref{third-eq-condi}. Also, the relation $\delta^{kn}\partial_{jkn}r_{AB}=-2\,n^j_{AB}/r^2_{AB}$ is utilized during the simplifications. 
Up to this point, as they have the same EMSG parameter, the internal and external dynamics appear to be governed by the same EMSG field. This is due to our assumption for the internal dynamics of the bodies, where the EMSG field is the same for all bodies, as well as the decomposition of the EMSG gravitational potentials into two internal and external parts with the same EMSG parameter. In the following, it is however shown that the final external dynamics of the body is independent of the EMSG terms.

Now, by collecting Eqs. \eqref{a-N}, \eqref{a-EMSG}, and \eqref{a-STR}, the result interestingly reduces to 
\begin{align}\label{a1}
\nonumber
&a^j_A[0\text{\tiny PN\normalsize}]+a_A^{\text{EMSG},j}[1\text{\tiny PN\normalsize}]+a^j_A[\text{STR}]~~~~~~~~~~~~~~~~\\
&~~~~~~~~~~~~~~~~~~~~~~~~~=-\sum_{B\neq A}G\,M_B\frac{n^j_{AB}}{r^2_{AB}},
\end{align}
where $M_B=m_B+\frac{E_B}{c^2}+O(c^{-4})$ is the total mass-energy of body $B$. The total energy defined in Eq. \eqref{total_energy} is used here with analogous expressions for body $B$. Also, in the GR fragment of the PN acceleration $a^{\text{GR},j}_A[1\text{\tiny PN\normalsize}]$, we can easily change the mass of each body, $m_i$ where $i=A,\,B,\,C$, to its total mass-energy, $M_i$, keeping the same form as given in Eq. \eqref{aPNGR}. This is because the difference between $m_i$ and $M_i$ induces a change in the order $c^{-4}$ of this relation, which is beyond the accuracy we focus on in the current study.  

Finally, the combination of these results describes the motion of a self-gravitating body among a system of $N$ bodies in the EMSG theory, which is
\begin{align}
\nonumber
&\bm{a}_A=-\sum_{B\neq A}G\,M_B\frac{n^j_{AB}}{r^2_{AB}}\\\nonumber
&-\frac{1}{c^2}\bigg\lbrace\sum_{B\neq A}\frac{G\,M_B}{r^2_{AB}}\Big[v_A^2-4\big(\bm{v}_A\cdot\bm{v}_B\big)+2v_B^2\\\nonumber
&-\frac{3}{2}\big(\bm{n}_{AB}\cdot\bm{v}_B\big)^2-\frac{5\,G\,M_A}{r_{AB}}-\frac{4\,G\,M_B}{r_{AB}}\Big]n^j_{AB}\\
&-\sum_{B\neq A} \frac{G\,M_B}{r^2_{AB}}\Big[\bm{n}_{AB}\cdot\big(4\bm{v}_A-3\bm{v}_B\big)\Big]\big(v^j_A-v^j_B\big)\\\nonumber
&-\sum_{B\neq A}\sum_{C\neq A,B}\frac{G^2 M_B\,M_C}{r^2_{AB}}\Big[\frac{4}{r_{AC}}+\frac{1}{r_{BC}}-\frac{r_{AB}}{2r^2_{BC}}\\\nonumber
&\times\big(\bm{n}_{AB}\cdot\bm{n}_{BC}\big)\Big]n^j_{AB}+\frac{7}{2}\sum_{B\neq A}\sum_{C\neq A,B}\frac{G^2M_B\,M_C}{r_{AB}r^2_{BC}}n^j_{BC}\bigg\rbrace.
\end{align}
It is important to notice that the final result does not depend on the internal structure of the bodies. Therefore, although this derivation relies on the assumption of perfect fluid bodies, the final equations of motion can well describe a system composed of solid bodies, such as Earth-like planets, and it can be applied for the solar-system and stellar-system experiments. Using this $N$-body acceleration, one can then examine the classical perihelion shift of the planets as well as the tests of GWEP in the EMSG theory. 
As seen, all the EMSG corrections are absorbed in the definition of the total mass-energy, while the extra EMSG contributions are canceled out well during the calculation so that the final form of the equations of motion is quite similar to those obtained in GR.  
Given that the total mass-energy is what is determined from the experiments, not the energy, this important finding shows that by studying the motion of an object in the weak-field limit of this theory, one cannot distinguish between EMSG and GR. A similar point has been raised in \cite{2023PDU....4201305A}. In other words, the EMSG theory passes the third Solar System test, the perihelion advance of Mercury, with flying colors. Moreover, as the motion of the massive self-gravitating body is structure-independent up to the 1\tiny PN \normalsize order, this theory does not violate GWEP.

\section{Inter-body metric}\label{secIV}
To complete our discussion in the previous section, we also examine the PN inter-body metric of the $N$-body system in the EMSG theory. This spacetime metric will govern the empty region between these well-separated bodies to the 1\tiny PN \normalsize order. 

Looking at Eqs. \eqref{G00}-\eqref{Gjk}, one realizes that for this purpose, the gravitational potentials $U$, $U^j$, $\Psi$, and $U_{\text{\tiny EMSG}}$ need to be obtained for an $N$-body system. We recall that $\Psi=2\Phi_1-\Phi_2+\Phi_3+4\Phi_4-\frac{1}{2}\Phi_5-\frac{1}{2}\Phi_6+f_0'c^4U_{\text{\tiny EMSG}}$ obtained in the previous section.  To do so, we first rewrite these potentials as $\phi(t,\boldsymbol{x})=\sum_A\int_A f(t,\boldsymbol{x}',\boldsymbol{x})d^3x'$ in which the form of the integrand $f(t,\boldsymbol{x}',\boldsymbol{x})$ is derived from the definitions given in \eqref{U}-\eqref{U_EMSG} and \eqref{Phi_1}-\eqref{Phi_6} for each gravitational potential. 
Next, the integration variables are changed to $\bar{\boldsymbol{x}}'=\boldsymbol{x}'-\boldsymbol{r}_A$ ( and to $\bar{\boldsymbol{v}}'=\boldsymbol{v}'-\boldsymbol{v}_A$ in cases where we have velocity). We then consider that $R_A\ll s_A$ in which $R_A$ is the characteristic size of body $A$ and $\boldsymbol{s}_A=\boldsymbol{x}-\boldsymbol{r}_A$. This assumption as well as the change of integration variable help us to find $|\boldsymbol{x}-\boldsymbol{x}'|^{-1}$ as the following Taylor expansion in powers of $\boldsymbol{x}'$:
\begin{align}
\frac{1}{|\boldsymbol{x}-\boldsymbol{x}'|}=\frac{1}{s_A}-\bar{x}'^j\partial_j\frac{1}{s_A}+\frac{1}{2}\bar{x}'^j\bar{x}'^k\partial_{jk}\frac{1}{s_A}+\cdots.
\end{align}  
Using this relation and invoking the assumptions mentioned at the beginning of Sec. \eqref{secIII}, after some calculations, we finally arrive at  
\begin{subequations}
\begin{align}
\label{eq1}
& U=\sum_{A}\frac{G\,m_A}{s_A}+\cdots,\\
\label{Uj_N}
& U^j=\sum_A\frac{G\,m_A\,v^j_A}{s_A}+\cdots,\\
\label{U_EMSG-N}
& U_{\text{\tiny EMSG}}=\sum_A\frac{G\,\mathfrak{M}_A}{s_A}+\cdots,\\
\nonumber
&\Psi=\sum_A\frac{G}{s_A}\Big(4\mathcal{T}_A+\frac{5}{2}\Omega_A+E^{\text{int}}_A+\frac{9}{2}P_A+\frac{3}{2}f_0'c^4\mathfrak{M}_A\Big)\\
\nonumber
&-\frac{1}{2}\sum_A\frac{G}{s_A}\Big(2T^{jk}_A+\Omega_A^{jk}+\delta^{jk}P_A+f_0'c^4\delta^{jk}\mathfrak{M}_A\Big)n_{Aj}n_{Ak}\\
&+\sum_A\frac{G\,m_A}{s_A}\Big(2v^2_A-\frac{1}{2}\big(\boldsymbol{n}_A\cdot{\boldsymbol{v}_A}\big)^2\Big)\\\nonumber
&-\sum_A\sum_{B\neq A}\frac{G^2\,m_A\,m_B}{r_{AB}s_A}\frac{5r^2_{AB}+s_A^2-s_B^2}{4r^2_{AB}}+\cdots,
\end{align}
\end{subequations}
where $s_A=|\boldsymbol{x}-\boldsymbol{r}_A|$ and $\boldsymbol{n}_A=\frac{\boldsymbol{s}_A}{s_A}$. Here, the ellipsis represents the terms of the order of magnitude $(R_A/s_A)^2$ and higher.
The last potential, $\Psi$, can be simplified further as  
\begin{align}\label{Psi}
\nonumber
&\Psi=\sum_A\frac{G}{s_A}\Big(E_A-4f_0'c^4\mathfrak{M}_A\Big)\\
\nonumber
&+\sum_A\frac{G\,m_A}{s_A}\Big(2v^2_A-\frac{1}{2}\big(\boldsymbol{n}_A\cdot{\boldsymbol{v}_A}\big)^2\Big)\\
&-\sum_A\sum_{B\neq A}\frac{G^2\,m_A\,m_B}{r_{AB}s_A}\frac{5r^2_{AB}+s_A^2-s_B^2}{4r^2_{AB}}.
\end{align} 
recalling the equilibrium conditions \eqref{first-eq-condi} and \eqref{third-eq-condi}.  We can now use these potentials to derive the PN metric of the $N$-body system.

For the time-time component of $g_{\mu\nu}$, adding the Newtonian potential \eqref{eq1} to the first term of the first sum in the above PN potential gives that $\frac{2}{c^2}\sum_A\frac{G\big(m_A+E_A/c^2\big)}{s_A}=\frac{2}{c^2}\sum_A\frac{G\,M_A}{s_A}$. Therefore, the structure term $E_A$ containing an EMSG portion is absorbed in the definition of the total mass-energy of body $A$.
Similar to our strategy in the previous section, we change all $m_A$s to $M_A$s in the PN potential $U^j$ \eqref{Uj_N} and also in the second and third sums of the PN potential $\Psi$ \eqref{Psi} with keeping the original form of these potentials. It should be noted that these changes practically make a difference of the 2\tiny PN \normalsize order, which is ignored here. 
Furthermore, the remaining terms in $g_{00}$ that involve the EMSG corrections, coming from the potential $\Psi$, i.e., the second term of the first sum of Eq. \eqref{Psi} and the potential $U_{\text{\tiny EMSG}}$ \eqref{U_EMSG-N}, all cancel each other out elegantly. These points lead to the conclusion that   
\begin{align}
&g_{00}=-1+\frac{2}{c^2}\sum_A\frac{G\,M_A}{s_A}+\frac{1}{c^4}\sum_A\frac{G\,M_A}{s_A}\Big(4v^2_A\\\nonumber
&-\big(\boldsymbol{n}_A\cdot\boldsymbol{v}_A\big)^2-2\frac{G\,M_A}{s_A}\Big)-\frac{1}{c^4}\sum_A\sum_{B\neq A}\frac{G^2\,M_A\,M_B}{s_A}\Big(\frac{2}{s_B}\\
\nonumber
&+\frac{5}{2\,r_{AB}}+\frac{s_A^2-s_B^2}{2\,r_{AB}^3}\Big)+O(c^{-6}),
\end{align}
In a similar manner, we also have
\begin{align}
&g_{0j}=-\frac{4}{c^3}\sum_A\frac{G\,M_A\,v_A^j}{s_A}+O(c^{-5}),\\
&g_{jk}=\Big(1+\frac{2}{c^2}\sum_A\frac{G\,M_A}{s_A}\Big)\delta_{jk}+O(c^{-4}).
\end{align}
As seen, there is no individual EMSG effect in the spacetime metric of the $N$-body system. Its form is the same as those presented in GR and only the mass-energy $M_A$, the center-of-mass position $\boldsymbol{r}_A$, and the velocity $\boldsymbol{v}_A$ of each body appear in this metric. This result is also compatible with GWEP in the EMSG theory.

\section{Conclusion}\label{secV}

The constraint of the new type of modified gravity theory named EMSG in the local scale would deepen our understanding of its properties and its power in solving the $\Lambda$CDM problems. The motion of a massive self-gravitating body in the $N$-body system may host the EMSG effects which in principle can be limited at the solar scale using the perihelion shift of the planets as well as experimental tests of SEP. The main focus of this work is to provide an appropriate $N$-body acceleration in the PN limit of the EMSG theory so that all possible EMSG effects are comprehensively extracted.  

To do so, we have introduced the PN hydrodynamic equations and center-of-mass variables in the EMSG theory. Then, defining the equilibrium conditions in this modified theory of gravity, the $N$-body acceleration as well as the PN inter-body metric of the $N$-body system in the quadratic-EMSG theory have been derived.
It has been shown that only the combination of the mass-energy $M_A$, the center-of-mass position $\boldsymbol{r}_A$, and the velocity $\boldsymbol{v}_A$ of each body appear in these relations. As these results are structure-independent, it has been concluded that GWEP is preserved in the PN limit of this theory.
The bottom line is that the structure-dependent effects related to the EMSG corrections have completely been encapsulated in the total mass-energy $M$ of each body and the rest have been canceled out well during the calculation, so that the apparent violation of GWEP in the EMSG theory has been settled down.
Moreover, it has been revealed that the mathematical form of the $N$-body acceleration and the PN inter-body metric of the $N$-body system is quite similar to those obtained in GR. It means that the EMSG theory passes the third Solar System test, the perihelion advance of Mercury, and the test of GWEP with flying colors. From another perspective, at this level of accuracy, it is not possible to constrain the free parameter of this theory and even distinguish it from GR using these local tests.  

As the final point, it should be mentioned that in the companion paper \cite{Nazari2023EMSG},
we will represent ``self-accelerations" of the body's center of mass in the EMSG theory studying general systems that are not necessarily reflection-symmetric. 

\section*{acknowledgments}

Helpful comments by the anonymous referee are gratefully acknowledged. 
We gratefully acknowledge useful discussions with Mahmood Roshan.
We would also like to thank Ivan De Martino and Nihan Kat{\i}rc{\i} for reading the first version of the manuscript and \"{O}zg\"{u}r Akarsu for providing constructive comments and introducing helpful references.
We are grateful to the Bernoulli Center at the EPFL campus in Lausanne for hospitality and support during the final stage of this work.

\bibliographystyle{apsrev4-1}
\bibliography{Classical_tests_of_EMSG}

\appendix

\section{Post-Newtonian metric of the perfect fluid in EMSG}\label{App_PN_limit}
In the Appendix A of \cite{2022PhRvD.105j4026N}, using the modern PN approach based on the harmonic gauge, the near-zone PN metric of a perfect fluid is approximately obtained in the EMSG theory. To do so, the iterative procedure is applied. In \cite{2023MNRAS.523.5452A}, it is pointed out that imposing the harmonic gauge conditions\footnote{We refer the reader unfamiliar with the modern PN approach where the harmonic gauge conditions are imposed to \cite{poisson2014gravity}.} at the last step of this procedure will provide us with the order of magnitude of the theory parameter. 
Following this point as well as using the result given in Appendix B of \cite{2022PhRvD.105j4026N}, we deduce that the term $c^2f_0'\rho^*$ must be at most of the order of the first PN corrections, i.e., $O(c^{-2})$.
Regarding this fact and evaluating each EMSG term in the components of the PN metric given in Eqs. (7a)-(7c) of \cite{2022PhRvD.105j4026N}, we truncate the results to $O(c^{-4})$, $O(c^{-3})$, $O(c^{-2})$, and $O(c^{-2})$ for $g_{00}$, $g_{0j}$, $g_{jk}$, and the determinant of the metric, respectively. Note that these orders give the metric to the 1\tiny PN \normalsize order which is sufficient for our calculation throughout the current study.      
The metric components of a perfect fluid in EMSG are then simplified as:
\begin{subequations}
\begin{align}
\nonumber
& g_{00}=-1+\frac{2}{c^2}U+\frac{2}{c^4}\Big(\Psi-U^2\Big)+8f'_0U_{\text{\tiny EMSG}}\\\label{G00}
&+O(c^{-6}),\\
& g_{0j}=-\frac{4}{c^3}U_j+O(c^{-5}),\\
\label{Gjk}
& g_{jk}=\delta_{jk}\Big(1+\frac{2}{c^2}U\Big)+O(c^{-4}),
\end{align}
\end{subequations}
and its determinant reduces to 
\begin{align}
\big(-g\big)=1+\frac{4}{c^2}U+O(c^{-4}).
\end{align}     
The integral form of the Newtonian potential $U$ and the PN gravitational potentials $U_j$, $U_{\text{\tiny EMSG}}$, and $\Psi$ are respectively given by
\begin{subequations}
\begin{align}
\label{U}
& U=G\int_{\mathcal{M}}\frac{{\rho^*}'}{\rvert{\boldsymbol{x}-\boldsymbol{x}'}\rvert}d^3x',\\
& U^j=G\int_{\mathcal{M}}\frac{{\rho^*}'v'^j}{\rvert{\boldsymbol{x}-\boldsymbol{x}'}\rvert}d^3x',\\
\label{U_EMSG}
& U_{\text{\tiny EMSG}}=G\int_{\mathcal{M}}\frac{{\rho^*}'^2}{\rvert\boldsymbol{x}-\boldsymbol{x}'\rvert}d^3x',\\
&\psi=G\int_{\mathcal{M}}\frac{{\rho^*}'}{{\rvert{\boldsymbol{x}-\boldsymbol{x}'}\rvert}}\Big(\Pi'+\frac{1}{2}v'^2-\frac{1}{2}U'\Big)d^3x',\\
& V=G\int_{\mathcal{M}}\frac{{\rho^*}'}{\rvert{\boldsymbol{x}-\boldsymbol{x}'}\rvert}\Big(\frac{3p'}{{\rho^*}'}+v'^2-\frac{1}{2}U'\Big)d^3x',\\
& X=G\int_{\mathcal{M}}{{\rho^*}'}\rvert{\boldsymbol{x}-\boldsymbol{x}'}\rvert d^3x'.
\end{align}
\end{subequations}
Note that $\Psi=\psi+V+\frac{1}{2}\partial_{tt} X$.
Here, the three-dimensional sphere $\mathcal{M}$ with the radius $\mathcal{R}$, which is of the order of a wavelength of the radiation, represents the near-zone region. Its boundary is denoted by $\partial \mathcal{M}$. 
It should be mentioned that in Eqs. (7a)-(7c) of \cite{2022PhRvD.105j4026N}, there are extra EMSG terms which should be removed from the first-order PN metric. In fact, these terms play a role in the higher PN orders.

For the calculations presented in Sec. \ref{PN_hyd}, the Christoffel symbols are also required. Considering the relation $\Gamma^{\mu}_{\alpha\beta}=\frac{1}{2}g^{\mu\nu}\Big(\partial_\alpha g_{\nu\beta}+\partial_\beta g_{\nu\alpha}-\partial_\nu g_{\alpha\beta}\Big)$, one can obtain 
them as  
\begin{subequations}
\begin{align}
&\Gamma^{0}_{00}=-\frac{1}{c^3}\partial_tU+O(c^{-5}),\\
&\Gamma^{0}_{0j}=-\frac{1}{c^2}\partial_jU+O(c^{-4}),\\
&\Gamma^{0}_{jk}=\frac{1}{c^3}\Big(\delta^{jk}\partial_tU+2\big(\partial_jU_k+\partial_kU_j\big)\Big)+O(c^{-5}),\\\nonumber
&\Gamma^{j}_{00}=-\frac{1}{c^2}\partial_jU+\frac{1}{c^4}\Big(4U\partial_jU-4\partial_tU_j-\partial_j\Psi\Big)\\
&-4f'_0\partial_j U_{\text{\tiny EMSG}}+O(c^{-6}),\\
&\Gamma^{j}_{0k}=\frac{1}{c^3}\Big(\delta^{jk}\partial_tU-2\big(\partial_kU_j-\partial_jU_k\big)\Big)+O(c^{-5}),\\
&\Gamma^{j}_{kn}=\frac{1}{c^2}\big(\delta_{jn}\partial_kU+\delta_{jk}\partial_nU-\delta_{kn}\partial_jU\big)+O(c^{-4}),
\end{align}
\end{subequations}
for the spacetime described by Eqs. \eqref{G00}-\eqref{Gjk}. We keep the above terms up to the orders that play a role in the 1\tiny PN \normalsize corrections to the energy-momentum conservation equations, cf. Eqs. \eqref{eq_E} and \eqref{s_com}.

\section{Definition of the new quantities}\label{app2}
In this appendix, the new scalar and tensorial quantities we encounter during the calculations are introduced. The first group is dedicated to the scalar ones:
\begin{subequations}
\begin{align}
\label{T_A}
&\mathcal{T}_A\equiv \frac{1}{2}\int_A \rho^*\bar{v}^2\,d^3x,\\
\label{Omega_A}
&\Omega_A\equiv -\frac{1}{2}G\int_A\frac{\rho^*{\rho^*}'}{\rvert{\boldsymbol{x}-\boldsymbol{x}'}\rvert}d^3x'd^3x,\\
\label{Q_A}
&Q_A\equiv\int_A\rho^*\bar{v}^k\partial_k\rho^*\,d^3x,\\
&P_A\equiv \int_Ap\,d^3x,\\
\label{mEMSG}
&\mathfrak{M}_A\equiv \int_A {\rho^*}^2 d^3x\\
\label{E_A}
&E_A^{\text{int}}\equiv \int_A \rho^*\Pi\,d^3x,\\
&H_A\equiv G\int_A\rho^*{\rho^*}'\frac{\bar{\boldsymbol{v}}'\cdot\big(\boldsymbol{x}-\boldsymbol{x}'\big)}{\lvert\boldsymbol{x}-\boldsymbol{x}'\rvert^3}d^3x'd^3x.
\end{align}
\end{subequations} 
Here, $\Pi=\epsilon/\rho^*$. The definitions \eqref{Q_A} and \eqref{mEMSG} are the new scalar quantities introduced in EMSG. According to Eq. \eqref{rho} and also the definition of the total time derivative $d/dt=\partial_t+v^k\partial_k$, it can be shown that there is a relation between these two terms such that $Q_A=\frac{1}{2}\frac{d\mathfrak{M}_A}{dt}$. We use this point to simplify the virial identity \eqref{dddotI}. 

The second group is devoted to the tensorial quantities:
\begin{subequations}
\begin{align}
\label{I^jk}
&I^{jk}_A\equiv \int_A\rho^*\bar{x}^j\bar{x}^k\,d^3x,\\
\label{S^jk}
&S^{jk}_A\equiv \int_A\rho^*\big(\bar{x}^j\bar{v}^k-\bar{x}^k\bar{v}^j\big)d^3x,\\
&\mathcal{T}^{jk}_A\equiv \frac{1}{2}\int_A\rho^*\bar{v}^j\bar{v}^k\,d^3x,\\
& L_A^{jk}\equiv \int_A\bar{v}^j\partial_k p\,d^3x,\\
&\Omega^{jk}_A\equiv -\frac{1}{2}G\int_A\rho^*{\rho^*}'\frac{\big(x-x'\big)^j\big(x-x'\big)^k}{\lvert\boldsymbol{x}-\boldsymbol{x}'\rvert^3}d^3x'd^3x,\\
&H^{jk}_A\equiv G\int_A\rho^*{\rho^*}'\frac{\bar{v}'^j\big(x-x'\big)^k}{\lvert\boldsymbol{x}-\boldsymbol{x}'\rvert^3}d^3x'd^3x,\\
&K^{jk}_A\equiv G\int_A\rho^*{\rho^*}'\frac{\bar{\boldsymbol{v}}'\cdot\big(\boldsymbol{x}-\boldsymbol{x}'\big)\big(x-x'\big)^j\big(x-x'\big)^k}{\lvert\boldsymbol{x}-\boldsymbol{x}'\rvert^5}d^3x'd^3x,\\
&Q^{jk}_A\equiv \int_A\rho^*\bar{v}^j\partial_k\rho^*\,d^3x,
\end{align}
\end{subequations}
among which $Q^{jk}_A$ is a new quantity defined in the framework of EMSG. Notice that $\mathcal{T}_A=\delta_{jk}\mathcal{T}^{jk}_A$, $\Omega_A=\delta_{jk}\Omega_A^{jk}$, $H_A=\delta_{jk}H^{jk}_A$, and $Q_A=\delta_{jk}Q^{jk}_A$.

\section{Force integrals}\label{app3}
The integral forces we deal with in Sec. \ref{secIII} are defined in this appendix. We first introduce their general forms and then attempt to simplify them regarding the assumptions mentioned in this section. In some of these forces, there is no explicit EMSG effect. So, we call them the GR forces. These forces are
\begin{subequations}
\begin{align}
\label{F0}
&F_0^j\equiv\int_A\big(\rho^*\partial_jU-\partial_jp\big)d^3x,\\
&F_1^j\equiv\frac{1}{2c^2}\int_Av^2\partial_jp\,d^3x,\\
&F_2^j\equiv\frac{1}{c^2}\int_AU\partial_jp\,d^3x,\\
&F_3^j\equiv\frac{1}{c^2}\int_A\Pi\partial_jp\,d^3x,\\
&F_4^j\equiv\frac{1}{c^2}\int_A\frac{p}{\rho^*}\partial_jp\,d^3x,\\
&F_5^j\equiv-\frac{1}{c^2}\int_Av^j\partial_tp\,d^3x,\\
&F_6^j\equiv\frac{1}{c^2}\int_A\rho^*v^2\partial_jU\,d^3x,\\
&F_7^j\equiv-\frac{4}{c^2}\int_A\rho^*U\partial_jUd^3x,\\
&F_8^j\equiv-\frac{3}{c^2}\int_A\rho^*v^j\partial_tU\,d^3x,\\
&F_9^j\equiv-\frac{4}{c^2}\int_A\rho^*v^jv^k\partial_kU\,d^3x,\\
&F_{10}^j\equiv\frac{4}{c^2}\int_A\rho^*v^k\partial_kU^j\,d^3x,\\
&F_{11}^j\equiv-\frac{4}{c^2}\int_A\rho^*v^k\partial_jU_k\,d^3x,\\
&F_{12}^j\equiv\frac{4}{c^2}\int_A\rho^*\partial_tU_j\,d^3x,\\
&F_{13}^j\equiv\frac{2}{c^2}\int_A\rho^*\partial_j\Phi_1\,d^3x,\\
&F_{14}^j\equiv-\frac{1}{c^2}\int_A\rho^*\partial_j\Phi_2\,d^3x,\\
&F_{15}^j\equiv\frac{1}{c^2}\int_A\rho^*\partial_j\Phi_3\,d^3x,\\
&F_{16}^j\equiv\frac{4}{c^2}\int_A\rho^*\partial_j\Phi_4\,d^3x,\\
&F_{17}^j\equiv-\frac{1}{2c^2}\int_A\rho^*\partial_j\Phi_5\,d^3x,\\
\label{F18}
&F_{18}^j\equiv-\frac{1}{2c^2}\int_A\rho^*\partial_j\Phi_6\,d^3x,
\end{align}
\end{subequations}
in which $\Phi_1$-$\Phi_6$ are called the auxiliary potentials given by the following relations
\begin{subequations}
\begin{align}
\label{Phi_1}
&\Phi_1\equiv G\int_{\mathcal{M}}\frac{{\rho^*}'v'^2}{\rvert{\boldsymbol{x}-\boldsymbol{x}'}\rvert}d^3x',\\
&\Phi_2\equiv G\int_{\mathcal{M}}\frac{{\rho^*}'U'}{\rvert{\boldsymbol{x}-\boldsymbol{x}'}\rvert}d^3x',\\
&\Phi_3\equiv G\int_{\mathcal{M}}\frac{{\rho^*}'\Pi'}{\rvert{\boldsymbol{x}-\boldsymbol{x}'}\rvert}d^3x',\\
&\Phi_4\equiv G\int_{\mathcal{M}}\frac{p'}{\rvert{\boldsymbol{x}-\boldsymbol{x}'}\rvert}d^3x',\\
&\Phi_5\equiv G\int_{\mathcal{M}}{\rho^*}'\partial_{j'}U'\frac{\big(x-x'\big)^j}{\rvert{\boldsymbol{x}-\boldsymbol{x}'}\rvert}d^3x',\\
\label{Phi_6}
&\Phi_6\equiv G\int_{\mathcal{M}}{\rho^*}'v'_jv'_k\frac{\big(x-x'\big)^j\big(x-x'\big)^k}{\rvert{\boldsymbol{x}-\boldsymbol{x}'}\rvert^3}d^3x'.
\end{align}
\end{subequations}

The other forces coming from the EMSG terms are
\begin{subequations}
\begin{align}
\label{F19}
&F_{19}^j\equiv 3f_0'c^2\int_A\rho^*v^2\partial_j\rho^*\,d^3x,\\
&F_{20}^j\equiv 14f_0'c^2\int_A \rho^*U\partial_j\rho^*\,d^3x,\\
&F_{21}^j\equiv -2f_0'c^2\int_A \rho^*\Pi \partial_j\rho^*\,d^3x,\\
&F_{22}^j\equiv 2f_0'c^2\int_A \rho^*v^jv^k\partial_k\rho^*\,d^3x,\\
&F_{23}^j\equiv 2f_0'c^2\int_A {\rho^*}^2v^j\partial_kv^k\,d^3x,\\
&F_{24}^j\equiv 2f_0'c^2\int_A \partial_j\big(\rho^*p\big)\,d^3x,\\
&F_{25}^j\equiv 4{f_0'}^2c^6\int_A {\rho^*}^2\partial_j\rho^*\,d^3x,\\
&F_{26}^j\equiv f_0'c^2\int_A {\rho^*}^2\partial_jv^2\,d^3x,\\
&F_{27}^j\equiv -2f_0'c^2\int_A {\rho^*}^2\partial_j\Pi\,d^3x,\\
&F_{28}^j\equiv 6f_0'c^2\int_A {\rho^*}^2\partial_jU\,d^3x,\\
&F_{29}^j\equiv 5f_0'c^2\int_A \rho^*\partial_j U_{\text{\tiny EMSG}}\,d^3x,\\
\label{F30}
&F_{30}^j\equiv -2f_0'c^4\int_A \rho^*\partial_j\rho^*\,d^3x.
\end{align}
\end{subequations}
Here, the last EMSG force, $F_{30}^j$, plays a role in the Newtonian limit, and it can be counted along with $F_0^j$ as the Newtonian contribution to the total force, while the rest, $F_{19}^j$-$F_{29}^j$, are considered as the EMSG forces acting on body $A$ in the first PN order. It seems that several EMSG forces may affect the motion of the body. In this work, we examine this point in detail and investigate the overall effects of these forces.  
 
We next obtain each force integral of $F_0^j\textendash F_{30}^j$ by applying the assumptions mentioned at the beginning of Sec. \ref{secIII}, rewriting every vector in terms of the corresponding relative vectors \eqref{rela-quan}, and also decomposing the Newtonian, PN, and EMSG gravitational potentials into two parts: the portion produced by body $A$, which is called the internal potential and is represented by
\begin{align}
\phi_A(t,\boldsymbol{x})=\int_Af(t,\boldsymbol{x}',\boldsymbol{x})d^3x',
\end{align}
and those produced by the remaining bodies in the $N$-body system, which are called the external potentials and are represented by
\begin{align}
\phi_{-A}(t,\boldsymbol{x})=\sum_{B\neq A}\int_Bf(t,\boldsymbol{x}',\boldsymbol{x})d^3x'.
\end{align}
Here, $\phi$ is an arbitrary gravitational potential that is obtained from the integral of the function $f(t,\boldsymbol{x}',\boldsymbol{x})$ over the volume occupied by the bodies of the considered system. The total potential for each case will therefore be  $\phi=\phi_A+\phi_{-A}$. For each gravitational potential, the form of the integrand $f(t,\boldsymbol{x}',\boldsymbol{x})$ is deduced from their definitions given in \eqref{U}-\eqref{U_EMSG} and \eqref{Phi_1}-\eqref{Phi_6}.

Another important point that comes from the well-separated bodies assumption and helps us to simplify these force integrals significantly is that when the external potential $\phi_{-A}$ should be evaluated within body $A$, it and its partial derivative can be written as the following Taylor expansions 
\begin{subequations}
\begin{align}
\nonumber
&\phi_{-A}(t,\boldsymbol{x})=\phi_{-A}(t,\boldsymbol{r}_A)+\bar{x}^j\partial_j\phi_{-A}(t,\boldsymbol{r}_A)\\
&~~~~~~~~~~~~~~~+\frac{1}{2}\bar{x}^j\bar{x}^k\partial_{jk}\phi_{-A}(t,\boldsymbol{r}_A)+\cdots,\\
&\partial_j\phi_{-A}(t,\boldsymbol{x})=\partial_j\phi_{-A}(t,\boldsymbol{r}_A)+\bar{x}^k\partial_{jk}\phi_{-A}(t,\boldsymbol{r}_A)+\cdots,
\end{align}
\end{subequations} 
where $\bar{\boldsymbol{x}}=\boldsymbol{x}-\boldsymbol{r}_{A}(t)$ is the position of a fluid element relative to the center-of-mass of body $A$, $\boldsymbol{r}_A(t)$.

The other useful integral identities utilized here are summarized below:
\begin{subequations}
\begin{align}
&\frac{d}{dt}\int f(t,\boldsymbol{x})d^3x=\int \frac{\partial f}{\partial t}d^3x,\\
&\frac{d}{dt}\int \rho^*(t,\boldsymbol{x})\,f(t,\boldsymbol{x})d^3x=\int\rho^*\frac{d f}{d t}d^3x,\\
&\frac{d}{dt}\int \rho^*(t,\boldsymbol{x}')\,f(t,\boldsymbol{x},\boldsymbol{x}')d^3x'=\int{\rho^*}'\Big(\frac{d f}{d t}\Big)_{\text{gen}}d^3x',\\
\nonumber
&\frac{d}{dt}\int \rho^*(t,\boldsymbol{x})\,\rho^*(t,\boldsymbol{x}')\,f(t,\boldsymbol{x},\boldsymbol{x}')d^3x'd^3x\\
&~~~~~~~~~~~~~~~~~~~=\int \rho^*{\rho^*}'\Big(\frac{d f}{d t}\Big)_{\text{gen}}d^3x'd^3x,\\
\nonumber
&\frac{\partial}{\partial t}\int \rho^*(t,\boldsymbol{x}')\,f(\boldsymbol{x},\boldsymbol{x}')d^3x'\\
&~~~~~~~~~~~~~~~~~~~=\int{\rho^*}'\boldsymbol{v}'\cdot\boldsymbol{\nabla}'f(\boldsymbol{x},\boldsymbol{x}')d^3x',
\end{align}
\end{subequations}
where $f$ is an arbitrary function that in fact plays the role of the fluid variables here, and 
\begin{subequations}
\begin{align}
&\frac{df}{dt}=\frac{\partial f}{\partial t}+\boldsymbol{v}\cdot\boldsymbol{\nabla}f,\\
& \Big(\frac{df}{dt}\Big)_{\text{gen}}=\frac{\partial f}{\partial t}+\boldsymbol{v}\cdot\boldsymbol{\nabla}f+\boldsymbol{v}'\cdot\boldsymbol{\nabla}'f,
\end{align}
\end{subequations}
are its usual and generalized total time derivatives, respectively. 

Utilizing the aforementioned points as well as Eq. \eqref{condi1}, after some algebra,  we find the forces:
\begin{subequations}
\begin{align}
\label{F0f}
&F_0^j=m_A\partial_jU_{-A},\\
&F_1^j=\frac{1}{c^2}v_A^kL_A^{kj},\\
&F_2^j=-\frac{1}{c^2}P_A\partial_jU_{-A},\\
&F_3^j=0,\\
&F_4^j=0,\\
&F_5^j=-\frac{1}{c^2}v_A^j\frac{d}{dt}P_A+\frac{1}{c^2}L_A^{jk}v^k_A,\\
&F_6^j=\frac{1}{c^2}\Big[2v_A^kH_A^{kj}+m_Av_A^2\partial_jU_{-A}+2\mathcal{T}_A\partial_jU_{-A}\Big],\\
&F_7^j=\frac{4}{c^2}\Big[2\Omega_A\partial_jU_{-A}-\Omega_{A}^{jk}\partial_kU_{-A}-m_AU_{-A}\partial_jU_{-A}\Big],\\
&F_8^j=\frac{3}{c^2}\Big[v_A^kH_A^{jk}-v_A^jH_A-m_Av_A^j\partial_tU_{-A}\Big],\\\nonumber
&F_9^j=-\frac{4}{c^2}\Big[v_A^jH_A+v_A^kH_A^{jk}+m_Av_A^jv_A^k\partial_kU_{-A}\\
&~~~~~~+2\mathcal{T}^{jk}_A\partial_kU_{-A}\Big],\\
&F_{10}^j=\frac{4}{c^2}\Big[v_A^jH_A-v^k_AH_A^{jk}+m_Av_A^k\partial_kU_{j,-A}\Big],\\
&F_{11}^j=-\frac{4}{c^2}m_Av_A^k\partial_jU_{k,-A},\\
&F_{12}^j=-\frac{4}{c^2}\Big[2\Omega_A\partial_jU_{-A}-v_A^kH_A^{jk}-v_A^jH_A-m_A\partial_tU_{j,-A}\Big],\\
&F_{13}^j=-\frac{2}{c^2}\Big[2v^k_AH^{kj}_A-m_A\partial_j\Phi_{1,-A}\Big],\\
&F_{14}^j=-\frac{1}{c^2}\Big[m_A\partial_j\Phi_{2,-A}-\Omega^{jk}_A\partial_kU_{-A}\Big],\\
&F_{15}^j=\frac{1}{c^2}m_A\partial_j\Phi_{3,-A},\\
&F_{16}^j=\frac{4}{c^2}m_A\partial_j\Phi_{4,-A},\\
&F_{17}^j=-\frac{1}{2c^2}\Big[2\Omega^{jk}_A\partial_kU_{-A}-2\Omega_A\partial_jU_{-A}+m_A\partial_j\Phi_{5,-A}\Big],\\
&F_{18}^j=-\frac{1}{c^2}\Big[v^j_AH_A+v_A^kH_A^{jk}-3v_A^kK_A^{jk}+\frac{1}{2}m_A\partial_j\Phi_{6,-A}\Big],
\end{align}
\end{subequations}

and
\begin{subequations}
\begin{align}
&F_{19}^j=6f_0'c^2v_A^kQ_A^{kj},\\
&F_{20}^j=-7f_0'c^2\mathfrak{M}_A\partial_jU_{-A},\\
&F_{21}^j=0,\\
&F_{22}^j=2f_0'c^2\Big[v_A^kQ_A^{jk}+v_A^jQ_A\Big],\\
&F_{23}^j=-4f_0'c^2v_A^jQ_A,\\
&F_{24}^j=0,\\
&F_{25}^j=0,\\
&F_{26}^j=-4f_0'c^2v_A^kQ_A^{kj},\\
&F_{27}^j=0,\\
&F_{28}^j=6f_0'c^2\mathfrak{M}_A\partial_jU_{-A},\\
&F_{29}^j=5f_0'c^2m_A\partial_jU_{\text{\tiny EMSG},-A}\\
\label{F30f}
&F_{30}^j=0.
\end{align}
\end{subequations}
In the above relations, all external potentials are evaluated at $ \boldsymbol{x}=\boldsymbol{r}_A (t)$  after differentiation. For simplicity, we also use the notation $\partial_j\phi_{-A}$ for $\partial_j\phi_{-A}(t,r_A)$.

\end{document}